\makeatletter
\declare@file@substitution{revtex4-1.cls}{revtex4-2.cls}
\makeatother

\documentclass[preprint]{aastex631}

\usepackage{natbib}
\usepackage{url}
\usepackage{xcolor}  %\textcolor{red}{}
\usepackage{array}
\begin{document}
\title{The Eclipsing $\delta$ Scuti Binary HIP 7666: Absolute Dimensions and Pulsation Properties}
%Running Head: HIP 7666: Dimensions and Pulsations
\correspondingauthor{Jae Woo Lee}
\email{jwlee@kasi.re.kr}
\author[0000-0002-5739-9804]{Jae Woo Lee}
\affil{Korea Astronomy and Space Science Institute, Daejeon 34055, Republic of Korea}

\author[0000-0002-8692-2588]{Kyeongsoo Hong}
\affil{Korea Astronomy and Space Science Institute, Daejeon 34055, Republic of Korea}

\author[0000-0002-8394-7237]{Min-Ji Jeong}
\affil{Korea Astronomy and Space Science Institute, Daejeon 34055, Republic of Korea}

\author[0000-0003-1916-9976]{Pakakaew Rittipruk}
\affil{National Astronomical Research Institute of Thailand, Chiang Mai 50200, Thailand}

\author[0000-0001-9339-4456]{Jang-Ho Park}
\affil{Korea Astronomy and Space Science Institute, Daejeon 34055, Republic of Korea}

\begin{abstract}
We present a comprehensive analysis of the pulsating eclipsing binary HIP 7666 (TIC 308953703), containing a $\delta$ Sct-type 
component, based on 120-s TESS photometry and 63 newly acquired high-resolution echelle spectra. While previous orbital solutions 
were limited by sparse spectroscopic sampling, our extensive radial velocity (RV) measurements enable a robust determination of 
the system’s physical parameters. The primary has an effective temperature of $T_{\rm eff,A}$ = 7376$\pm$72 K and 
a projected rotational velocity of $v_{\rm A}\sin i$ = 56.6$\pm$5.3 km s$^{-1}$. Joint modeling of the light and RV curves shows 
that HIP 7666 is a detached system in a circular orbit, with Roche-lobe filling factors of $f_{\rm A}$ = 57 \% and 
$f_{\rm B}$ = 44 \%. The masses and radii are determined to better than 2 \% precision: $M_{\rm A}$ = 1.822$\pm$0.029 $M_\odot$, 
$R_{\rm A}$ = 2.154$\pm$0.017 $R_\odot$ for the primary, and $M_{\rm B}$ = 1.246$\pm$0.020 $M_\odot$, $R_{\rm B}$ = 1.170$\pm$0.015 $R_\odot$ 
for the secondary. The binary-based distance (316$\pm$8 pc) agrees well with the Gaia DR3 parallax-based distance (313$\pm$1 pc), 
validating the derived system parameters. Frequency analysis of the light-curve residuals reveals 15 significant signals, 
predominantly in the $\delta$ Sct pulsation frequency range. PARSEC evolutionary tracks indicate an age of $\sim$0.9 Gyr and 
a supersolar metallicity ($Z=0.03$), with the primary on the main sequence and the secondary located close to the zero-age main sequence. 
Galactic kinematics place HIP 7666 in the thin disk. 
\end{abstract}

%\keywords{Eclipsing binary stars (444) --- Spectroscopic binary stars (1557) --- Fundamental parameters of stars (555) 
%--- Stellar pulsations (1625) --- Delta Scuti variable stars (370) --- Stellar evolution (1599)}

\section{Introduction} \label{sec_intro}

Precise measurements of fundamental stellar parameters ($M$, $R$, $T_{\rm eff}$) serve as essential benchmarks for calibrating and 
refining models of stellar structure and evolution. Eclipsing binaries (EBs) provide one of the most reliable means of determining 
these properties across a broad range of evolutionary stages. In particular, detached main-sequence eclipsing binaries (dEBs) 
enable the most direct and nearly model-independent determination of absolute dimensions. Since their components evolve without 
significant mass transfer, dEBs allow robust measurements of masses and radii when full light curves are combined with 
well-sampled radial velocities (RVs) from both components. Such analyses can achieve precisions better than 3 \% \citep{Andersen1991,Torres+2010}, 
and recent advances in space-based photometry and high-resolution echelle spectroscopy have reduced uncertainties to 
the $\sim$0.2 \% level in favorable cases \citep[e.g.,][]{Maxted+2020,Southworth+2026}. These benchmark systems therefore place 
stringent constraints on current stellar models. 

The synergy between binary dynamics and asteroseismology offers an even more powerful probe of stellar interiors. 
Pulsating eclipsing binaries (PEBs) constitute a unique laboratory for stellar astrophysics. Within the instability strip of 
the Hertzsprung$-$Russell (H-R) diagram, intermediate-mass A$-$F type stars exhibit two principal classes of pulsations: 
$\delta$ Sct and $\gamma$ Dor variables. $\delta$ Sct stars pulsate in low-order pressure modes ($p$-modes), driven by 
the $\kappa$ mechanism operating in the He~II ionization zone \citep{Pamyatnykh1999,Aerts+2010}. In contrast, $\gamma$ Dor stars 
oscillate in high-order gravity modes ($g$-modes), which are excited by the convective flux blocking mechanism at the base 
of their envelope convection zones \citep{Guzik+2000}. The frequency boundary between these two classes is typically found 
near 4--5 day$^{-1}$, and stars exhibiting both types of modes are referred to as $\delta$ Sct--$\gamma$ Dor hybrids 
\citep[e.g.,][]{Grigahcene+2010, Uytterhoeven+2011}. These pulsation modes probe complementary regions of the stellar interior, 
ranging from the outer envelope to near the core. When such pulsators reside in EB systems, independently measured masses and 
radii from binary dynamics significantly reduce the degeneracies in asteroseismic modeling \citep{Murphy2018,Southworth+2026}. 
High-precision measurements of these systems also enable quantitative investigations of binarity and tidal effects on 
pulsation properties \citep{Soydugan+2006,Kahraman+2017}. 

Despite the availability of ultra-precise, high-cadence light curves from Kepler and TESS, many PEBs remain poorly characterized 
due to a lack of high-quality multi-epoch spectroscopy. Time-series observations at high spectral resolution are essential for 
determining accurate orbital solutions, atmospheric parameters, and absolute dimensions. To address this issue, we have initiated 
a long-term spectroscopic monitoring program of both candidate and known PEBs using 2 m-class telescopes equipped with 
echelle spectrographs. In this paper, we analyze HIP 7666 (TIC 308953703, TYC 3671-1094-1, Gaia DR3 407170600540598144; $T_{\rm p}$ 
= $+$9.242; $V_{\rm T}$ = $+$9.688, $(B-V)_{\rm T}$ = $+$0.458), a dEB exhibiting $\delta$ Sct-type pulsations. 

A thorough review of the research history of HIP 7666 is provided by \citet{Feng+2021}; here, we summarize only the results most 
relevant to its physical parameters and pulsation characteristics. The $\delta$ Sct-type variability of this system was first 
reported by \citet{Escola-Sirisi+2005}, who tentatively identified a frequency signal at either 24.46 day$^{-1}$ or 25.47 day$^{-1}$. 
Subsequently, \citet{Feng+2021} derived the first set of absolute dimensions by combining ground-based $BVR$ photometry with 
low- and high-resolution spectroscopy, assuming a circular orbit ($e = 0$). From the binary-subtracted residuals, they extracted 
three pulsation frequencies: $f_1$ = 24.631 day$^{-1}$, $f_2$ = 21.193 day$^{-1}$, and $f_3$ = 28.07 day$^{-1}$. 
Most recently, \citet{Pawar+2024} reanalyzed the binary target using TESS Sector 58 data (120-s cadence) alongside the RV measurements 
from \citet{Feng+2021}. While they identified seven significant frequencies and proposed a slightly eccentric orbit ($e = 0.0270$) 
based on the previously published RVs, their derived fundamental parameters remain largely consistent with those of the earlier study, 
as both analyses relied on the same limited spectroscopic dataset. 

Notably, the RV solutions in these previous works were based on only 10 spectroscopic epochs, resulting in insufficient 
orbital phase coverage. Such sparse sampling introduces significant uncertainties in the determination of the spectroscopic 
orbital elements (systemic velocity, RV semi-amplitudes, mass ratio), and may consequently bias the derived physical dimensions. 
To overcome these limitations, we have secured a significantly expanded dataset comprising 63 high-resolution echelle spectra 
with a relatively uniform phase distribution. By combining this extensive spectroscopic dataset with TESS photometry, 
we aim to refine the absolute dimensions of HIP 7666, reassess the previously reported pulsation frequencies, and provide 
a more robust characterization of the system’s binary dynamics and evolutionary status.

\section{TESS Photometry and Eclipse Timings} \label{sec_photometry}

The continuous space-based light curve of HIP 7666 was obtained from Sector 58 (S58) of the TESS mission \citep{Ricker+2015}, 
spanning the period from 2022 October 29 to November 26 (BJD 2,459,882.33 $-$ 2,459,910.06). We employed the 120-s cadence data, 
which were reduced by the SPOC pipeline \citep{Jenkins+2016} and retrieved from the MAST portal\footnote{\url{https://mast.stsci.edu/portal/Mashup/Clients/Mast/Portal.html}. The corresponding data can be accessed via \url{http://dx.doi.org/10.17909/d6tf-9118}.}. 
The CROWDSAP value of 0.961 indicates that approximately 3.9 \% of the flux within the photometric aperture is attributed to 
nearby contaminating sources. For our analysis, the SAP fluxes were detrended and converted into magnitude units using 
a maximum-light value of $T_{\rm p}$ = $+$9.242 \citep{Paegert+2022}. A detailed description of this reduction process is 
provided in \citet{Lee+2019,Lee+2024}. Figure~\ref{Fig1} illustrates the time-series (top) and phase-folded (middle) light curves 
for the binary target, comprising a total of 18,631 individual measurements, represented by black circles. 

From the processed data, we derived 23 mid-eclipse timings—11 for the primary (Min I) and 12 for the secondary (Min II) eclipses— 
via the \citet{Kwee+1956} method. These values, along with their associated uncertainties, are compiled in Table~\ref{Tab1}. 
By combining our eclipse timings with the available historical measurements \citep{Escola-Sirisi+2005,Feng+2021}, we derived 
the following linear ephemeris for HIP 7666 using an ordinary least-squares fit: 
\begin{equation}
C = \mbox{BJD}~ 2,451,769.56299(50) + 2.37222188(18) \times E. 
\end{equation}
Since the literature epochs were originally reported in HJD, we converted them into BJD$_{\text{TDB}}$ using the online 
utility\footnote{\url{https://arbiter.nextastro.org/toolkit/bjd-converter}} provided by \citet{Eastman+2010} to ensure consistency 
with our TESS observations. The $O-C$ timing residuals computed from this ephemeris are plotted in the upper panel of 
Figure~\ref{Fig2}. Due to the high density and overlap of the precise TESS timings, a magnified view of the $O-C$ diagram 
focusing solely on these space-based observations is presented in the lower panel. As shown, the two sets of minima partially 
overlap; however, the secondary minima exhibit a mean offset of approximately 0.00019 days relative to the primary ones. 
This small offset may be associated with the pulsations of the primary star (discussed in Section~\ref{sec_pulsation}) rather 
than with apsidal motion in an eccentric orbit.

\section{Echelle Spectroscopy and Data Analysis} \label{sec_spectroscopy}

Spectroscopic observations of HIP 7666 were conducted using the BOES fiber-fed by the 1.8-m telescope at the Bohyunsan Optical 
Astronomy Observatory (BOAO), and the MRES coupled via fiber to the 2.4-m telescope at the Thai National Observatory (TNO). 
Detailed descriptions of these instruments are provided by \citet{Kim+2007} for BOES, and by \citet{Buisset+2018} and \citet{Errmann+2020} 
for MRES. We adopted the observation and data reduction protocols established by \citet{Lee+2026} for TIC 399725538. 
At BOAO, we obtained 60 high-resolution spectra ($R \approx 30,000$) over 13 nights from 2018 February to 2021 February. 
Supplemental observations were carried out at TNO on 2019 November 30, yielding three additional spectra with $R \approx 18,000$. 
Both facilities utilized a consistent exposure time of 1800 s, achieving a typical signal-to-noise ratio (S/N) of $\sim$40 at 
4500 \AA. The reduction pipeline, implemented using the \texttt{IRAF}/\texttt{echelle} package, encompassed standard procedures: 
bias subtraction, flat-fielding, Th–Ar lamp-based wavelength calibration, order extraction, and continuum normalization.

The RVs of HIP 7666 were extracted via the broadening function (BF) technique \citep{Rucinski2002}, as executed within 
\texttt{RaveSpan} \citep{Pilecki+2012}. A synthetic LTE spectrum from the library of \citet{Coelho+2005} served as 
the reference template. The analysis focused on the spectral range of 4700–5450 \AA, which is largely free of telluric 
features and contains prominent hydrogen and metallic lines, most notably H$\beta$ and the Mg \textsc{i} b triplet. 
Representative BF profiles at BJD 2,459,169.1576 ($\phi=0.267$) and BJD 2,458,817.0302 ($\phi=0.830$) are displayed in 
Figure~\ref{Fig3}. The profiles exhibit distinct double-peaked structures, clearly resolving the two stellar components. 
They were fitted with rotational broadening functions to determine the RVs, which are tabulated in Table~\ref{Tab2}. 
The phase-folded RV curves are presented in Figure~\ref{Fig4}, tracing the orbital motions of both components. 

To derive the atmospheric parameters, we first disentangled the observed spectra using the \texttt{DISENTANGLING\_SHIFT\_AND\_ADD} 
code \citep{Shenar+2020, Shenar+2022}. This process was performed over the wavelength range of 3800–5080 \AA, which covers 
several temperature-sensitive spectral features, including the Balmer series (H$_9$ to H$\beta$) and the Ca \textsc{ii} K and 
H lines (3934 \AA\ and 3968 \AA, respectively). The initial orbital parameters (period $P_{\rm orb}$, reference epoch $T_0$, 
eccentricity $e$, and RV semi-amplitudes $K_{\rm A,B}$) required for the procedure were adopted from the binary modeling results 
presented in the following section. HIP 7666 is a dEB system in which both components contribute substantially to the total flux 
and undergo partial eclipses. As such, spectra obtained near orbital phases 0.0 and 0.5 are heavily blended, which can compromise 
the reliability of the spectral disentangling. To mitigate potential artifacts, we performed the disentangling 
using only spectra acquired within the phase ranges of 0.1–0.4 and 0.6–0.9. Despite this approach, the reconstructed spectrum 
of the secondary star exhibited significant noise due to its lower flux contribution. Consequently, the secondary's spectrum 
was deemed insufficient for a reliable atmospheric analysis, and our subsequent parameter determination focused exclusively on 
the primary component. 

The primary component's disentangled spectrum achieved an S/N of approximately 55 in the 4400–4800 \AA \, range. This spectrum 
was first continuum-normalized via \texttt{SUPPNET} \citep{Rozanski+2022} and its atmospheric parameters were then derived 
using the \texttt{GSSP} package \citep{Tkachenko2015}. The underlying synthetic grid was computed using the \texttt{SynthV} code 
\citep{Tsymbal1996} in combination with LLmodels stellar atmospheres \citep{Shulyak+2004}. For the analysis, 
the initial metallicity was assumed to be the solar value; however, it was subsequently fixed at [M/H] = +0.3, a value consistent 
with the evolutionary models (Section~\ref{sec_conclusion}) given the fundamental parameters obtained from binary modeling. 
Additionally, the surface gravity was fixed at $\log g_{\rm A}$ = 4.0, as constrained by the dynamical parameters (Section~\ref{sec_binarity}). 
Based on these settings, the effective temperature ($T_{\rm eff,A}$) and projected rotational 
velocity ($v_{\rm A}\sin i$) were explored in a \texttt{GSSP} grid search over the ranges 6000$-$8000 K and 20$-$70 km s$^{-1}$, 
respectively. 

The final results yielded $T_{\rm eff,A}$ = 7376$\pm$72 K and $v_{\rm A}\sin i$ = 56.6$\pm$5.3 km s$^{-1}$. These values were 
derived from a polynomial fit to the minimum $\chi^2$ distribution, following the approach of \citet{Lehmann+2011}. 
The disentangled spectrum of the primary, overlaid with its best-fitting synthetic model, is presented in Figure~\ref{Fig5}. 
The microturbulent and macroturbulent velocities were not treated as free parameters in the \texttt{GSSP} grid search. Instead, 
they were estimated a posteriori with the \texttt{iSpec} package\footnote{\url{https://www.blancocuaresma.com/s/iSpec/}} 
\citep{Blanco-Cuaresma+2014} based on the best-fitting atmospheric parameters, yielding $\xi_{\rm mic,A}$ = 2.06 km s$^{-1}$ 
and $\zeta_{\rm mac,A}$ = 19.71 km s$^{-1}$, respectively.

\section{Binary Modeling and Absolute Parameters} \label{sec_binarity}

\begin{figure}
\includegraphics[scale=1.0]{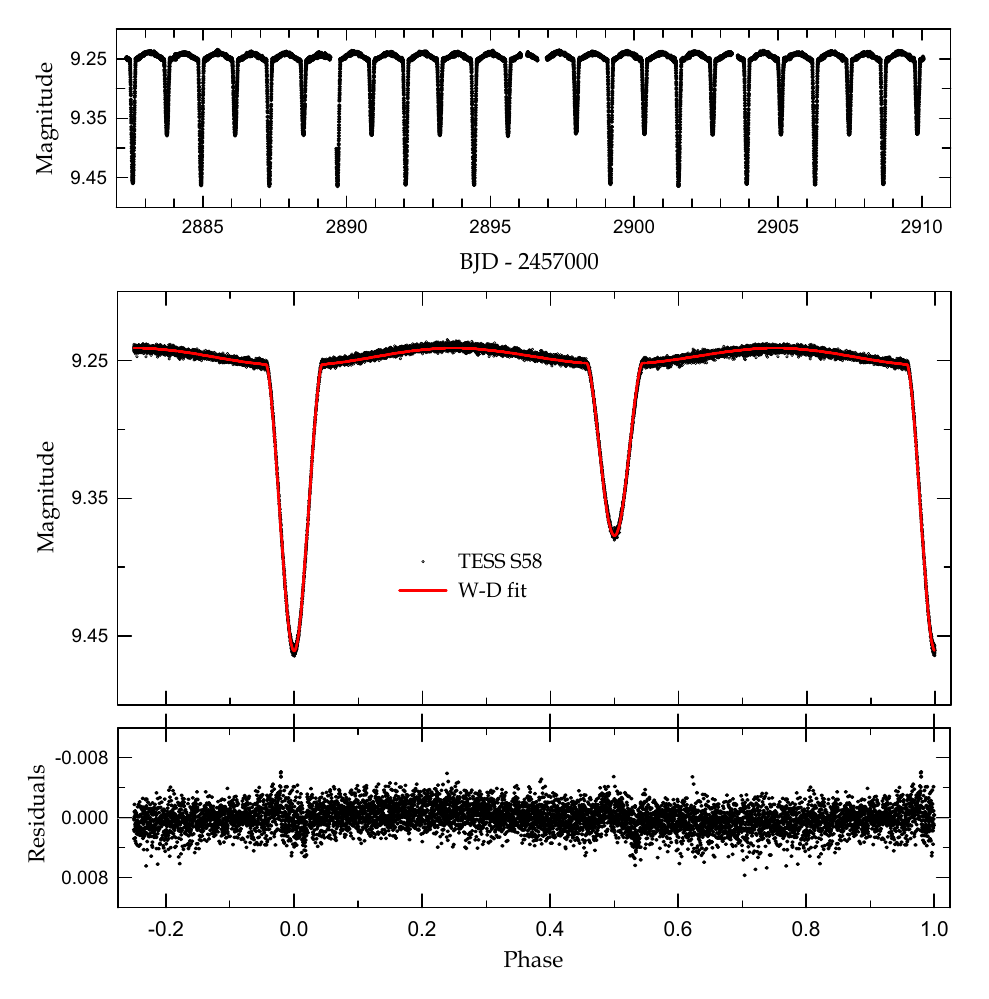}
\caption{TESS observations of HIP 7666 from Sector 58, plotted against BJD (top panel) and orbital phase (middle panel). 
Black circles represent individual measurements, while the red solid line shows the synthetic curve computed from 
the binary modeling parameters in Table~\ref{Tab3}. The corresponding residuals are displayed in the bottom panel. }
\label{Fig1}
\end{figure}

\begin{figure}
\includegraphics[]{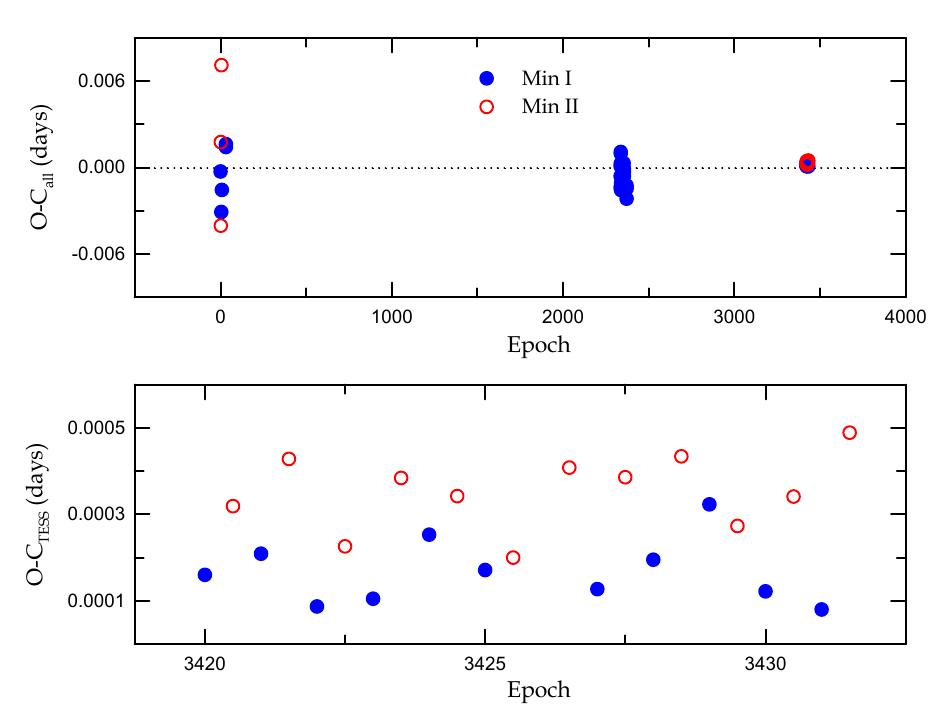}
\caption{Historical eclipse timing $O$--$C$ residuals for HIP 7666 constructed with the orbital ephemeris (1). The filled 
and open circles represent the times of the primary and secondary eclipses, labeled Min I and Min II, respectively. 
The lower panel shows only the precise TESS mid-eclipse times obtained in Sector 58 (Oct/Nov 2022; see Table~\ref{Tab1}).} 
\label{Fig2}
\end{figure}

\begin{figure}
\includegraphics{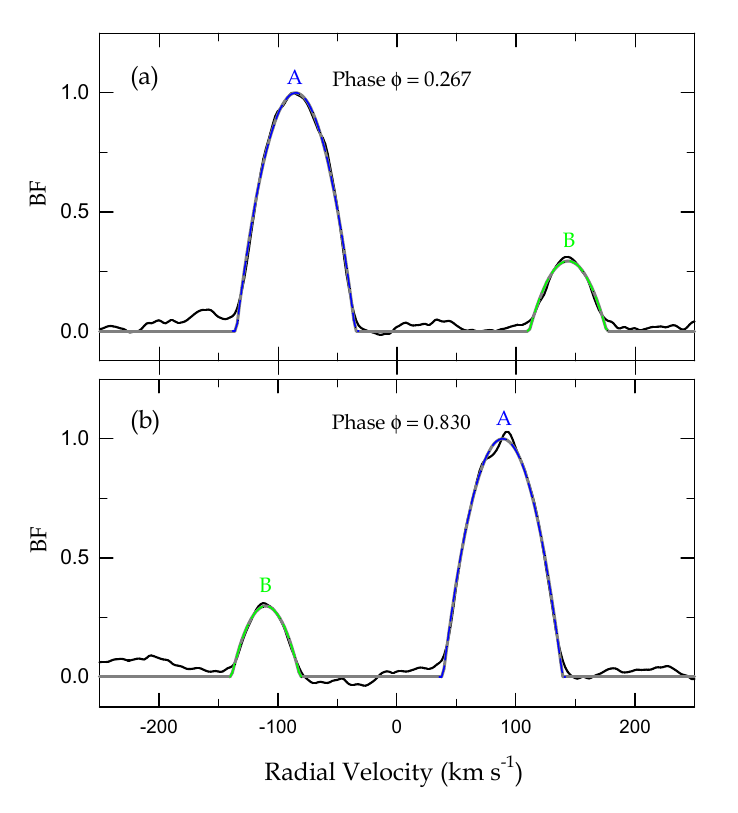}
\caption{Normalized BF profiles at two orbital phases $\phi$ = 0.267 (upper panel) and $\phi$ = 0.830 (lower panel), obtained 
using the \texttt{RaveSpan} code. The black solid lines represent the observed BF profiles, which exhibit a double-peaked structure 
originating from the two components. The blue and green dashed lines denote the best-ﬁt rotational broadening proﬁles for 
the primary (A) and secondary (B) components, respectively, while the gray lines correspond to the combined fits of these profiles. } 
\label{Fig3}
\end{figure}

As illustrated in Figure~\ref{Fig1}, the TESS light curve of HIP 7666 features a characteristic detached binary morphology. 
The primary and secondary eclipse depths are 0.211$\pm$0.011 and 0.131$\pm$0.007 mag, respectively. The depth discrepancy of 
0.080 mag suggests a distinct temperature difference between the stellar components. Furthermore, while the first maximum 
(Max I) appears slightly brighter than the second (Max II) by 0.0015$\pm$0.0023 mag, this difference is statistically insignificant 
within the measured uncertainties. Consequently, the O'Connell effect is considered negligible in the current TESS dataset. 

To obtain an optimal binary solution from the observables and refine the physical parameters of the target, we employed 
the Wilson-Devinney (W-D) binary program \citep{Wilson+1971, Kallrath2022}, which allows for the simultaneous modeling of 
light and RV curves via a differential correction algorithm. Our modeling approach parallels the analysis previously 
performed on TIC 322208686 \citep{Lee+2026}. Based on the analysis of the disentangled spectrum, the surface temperature of 
the primary star (HIP 7666 A) was fixed at $T_{\rm eff,A}$ = 7376$\pm$72 K throughout the modeling process, and we accounted 
for the propagated uncertainties originating from the error in $T_{\rm eff,A}$. The rotational parameter of the primary star, 
derived from the observed $v_{\rm A} \sin i$ and the calculated synchronous velocity ($v_{\rm sync} \sin i = 2 \pi R_{\rm A} \sin i / P_{\rm orb}$), 
was determined to be $F_{\rm A}$ = 1.25$\pm$0.12. As this value lies near the $2\sigma$ boundary of synchronous rotation, and 
its impact on the overall binary modeling is negligible, we adopted synchronous rotation ($F_{\rm A} = F_{\rm B} = 1.0$) 
for the subsequent analysis. 

Based on their effective temperatures, we treated the primary and secondary components as having radiative and 
convective envelopes, respectively. For the primary, the bolometric albedo and gravity-darkening coefficient were set to
$A_{\rm A} = 1.0$ and $g_{\rm A} = 1.0$, respectively \citep{Rucinski1969, vonZeipel1924}. For the cooler secondary, we adopted 
$A_{\rm B} = 0.5$ and $g_{\rm B} = 0.32$ \citep{Rucinski1969, Lucy1967}. A non-linear logarithmic limb-darkening (LD) law was 
employed, with initial coefficients ($x_{\rm A,B}$, $y_{\rm A,B}$) sourced from the updated tables of \citet{van1993} and 
subsequently adjusted during the W-D synthesis. In addition to the LD coefficients, our modeling involved the optimization of 
eleven further parameters: the spectroscopic parameters ($a$, $\gamma$, $q$), the geometric and stellar parameters ($i$, $T_{\rm B}$, 
$\Omega_{\rm A,B}$, $L_{\rm A}$, $l_3$), and the ephemeris parameters ($T_0$, $P_{\rm orb}$). 

\begin{figure}
\includegraphics[angle=0]{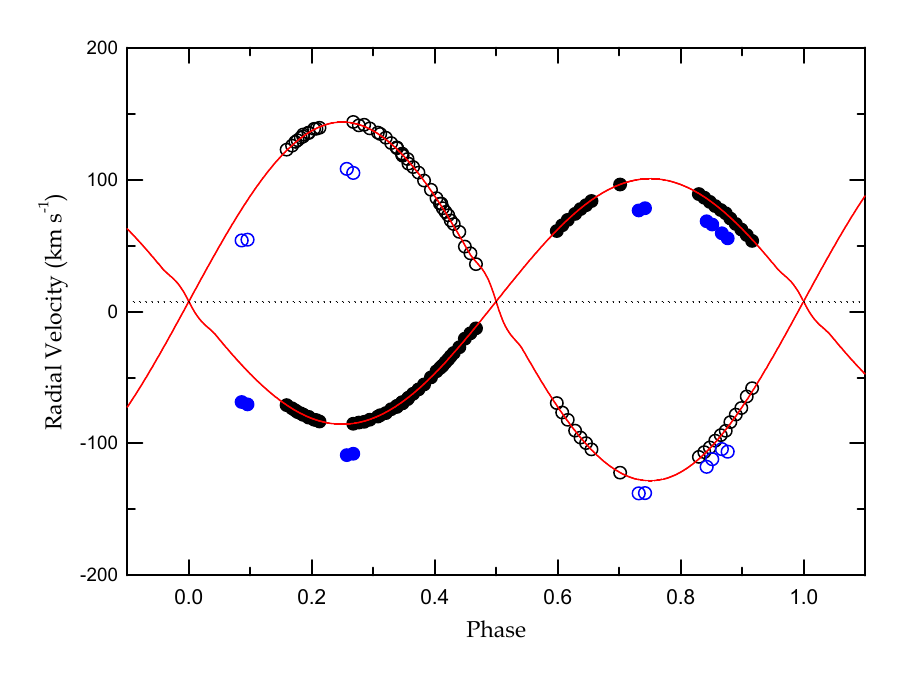}
\caption{RV curves of HIP 7666 with the best-fit models. The blue and black circles represent the RV measurements from 
\citet{Feng+2021} and this work, respectively. The red solid curves denote the W-D model fit to our double-lined RVs, 
while the dotted gray line indicates the systemic velocity of $\gamma$ = $+$7.64 km s$^{-1}$. } 
\label{Fig4}
\end{figure}

\begin{figure}
\includegraphics[scale=0.9]{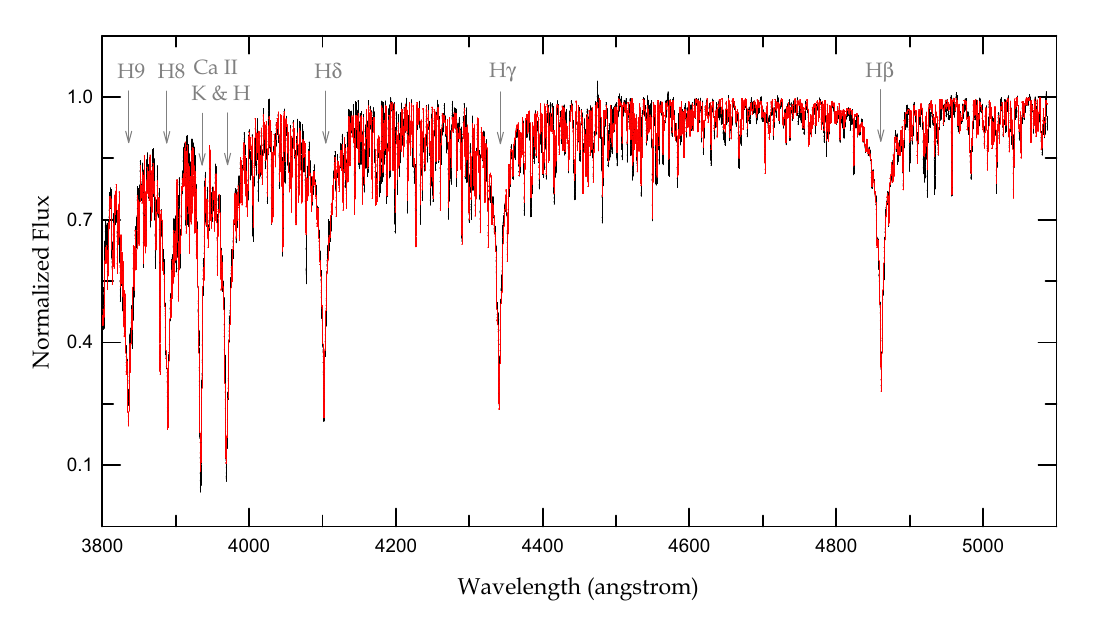}
\caption{Disentangled spectrum of HIP 7666 A. The black and red solid lines represent the disentangled and best-ﬁtting 
synthetic spectra, respectively. }
\label{Fig5}
\end{figure}

The final parameters derived from the binary synthesis are summarized in Table~\ref{Tab3}, and the corresponding best-fit models 
for the light and RV curves are illustrated in Figures~\ref{Fig1} and \ref{Fig4}, respectively. We also tested for 
potential orbital eccentricity ($e$) and found no significant deviation from a circular orbit. An independent Keplerian fit 
to our RV measurements yielded $e=0.003\pm0.002$, supporting the adoption of a circular orbit, in contrast to $e=0.0270\pm0.0009$ 
reported by \citet{Pawar+2024}. Our photometric-spectroscopic parameters in Table ~\ref{Tab3} show notable discrepancies with 
the results of \citet{Feng+2021}, which were based on their own double-lined RVs and ground-based $BVR$ photometry. Similarly, 
our findings differ from those of \citet{Pawar+2024}, who analyzed the same TESS S58 data used in our study but incorporated 
the RVs from \citet{Feng+2021}. Specifically, our mass ratio ($q$ = 0.684) is considerably lower than the values of 0.803 and 
0.780 reported by \citet{Feng+2021} and \citet{Pawar+2024}, respectively. As illustrated in Figure~\ref{Fig4}, 
this discrepancy reflects the marked differences between their RV data and ours. Given the limited number of observations and 
sparse phase coverage in the \citet{Feng+2021} dataset, we suggest that our more extensive RV dataset provides stronger constraints 
on the binary parameters. Moreover, the RV measurements obtained independently with BOES and MRES show excellent agreement, 
demonstrating the internal consistency of our spectroscopic data. 

The absolute dimensions of HIP 7666, derived from our preferred binary model, are summarized in Table~\ref{Tab4}. 
These values were determined by adopting solar standards of $T_{\rm eff,\odot} = 5780$ K and $M_{\rm bol,\odot} = +4.73$, with 
bolometric corrections (BCs) incorporated via an empirical relation as a function of $T_{\rm eff}$ \citep{Flower1996,Torres2010}. 
For the distance determination of HIP 7666 AB, we utilized the Johnson magnitudes of $V_{\rm system}$ = +9.65$\pm$0.03 and 
($B-V$) = +0.39$\pm$0.03, transformed from the Tycho-2 catalog \citep{Hog+2000}. Based on the color$-$temperature relation of 
\citet{Flower1996}, the primary temperature ($T_{\rm eff,A}$) yields an intrinsic color of ($B-V$)$\rm_0$ = +0.27$\pm$0.01, 
resulting in an interstellar reddening of $E$($B-V$) = +0.12$\pm$0.03. Accounting for an extinction of 
$A_{\rm V}$ $\simeq$ 3.1$E(B-V)$ and a third-light contribution of $l_3$ = 0.0375, we derived an EB-based distance of 
$316 \pm 8$ pc. Notably, this $l_3$ value is consistent with the TESS contamination level of $\sim$3.9 \% mentioned 
in Section~\ref{sec_photometry}, which further supports the reliability of our light curve synthesis. 
Our distance result is also in excellent agreement with the distance of $313 \pm 1$ pc inferred from the Gaia DR3 parallax 
($\pi = 3.191 \pm 0.013$ mas; \citealt{Gaia2023}). This agreement reinforces the robustness of our binary parameters.

\section{Pulsational Characteristics} \label{sec_pulsation}

HIP 7666 is a well-known EB system exhibiting $\delta$ Sct-type pulsations \citep{Escola-Sirisi+2005,Feng+2021,Pawar+2024}. 
The improved fundamental parameters presented in Table~\ref{Tab4}, derived from our echelle time-series spectroscopy, 
indicate that the primary component is a strong candidate for the $\delta$ Sct variable in this system, consistent with 
the typical characteristics of such pulsators \citep{Breger2000, Antoci+2019}. 

\begin{figure}
\includegraphics[scale=1.0]{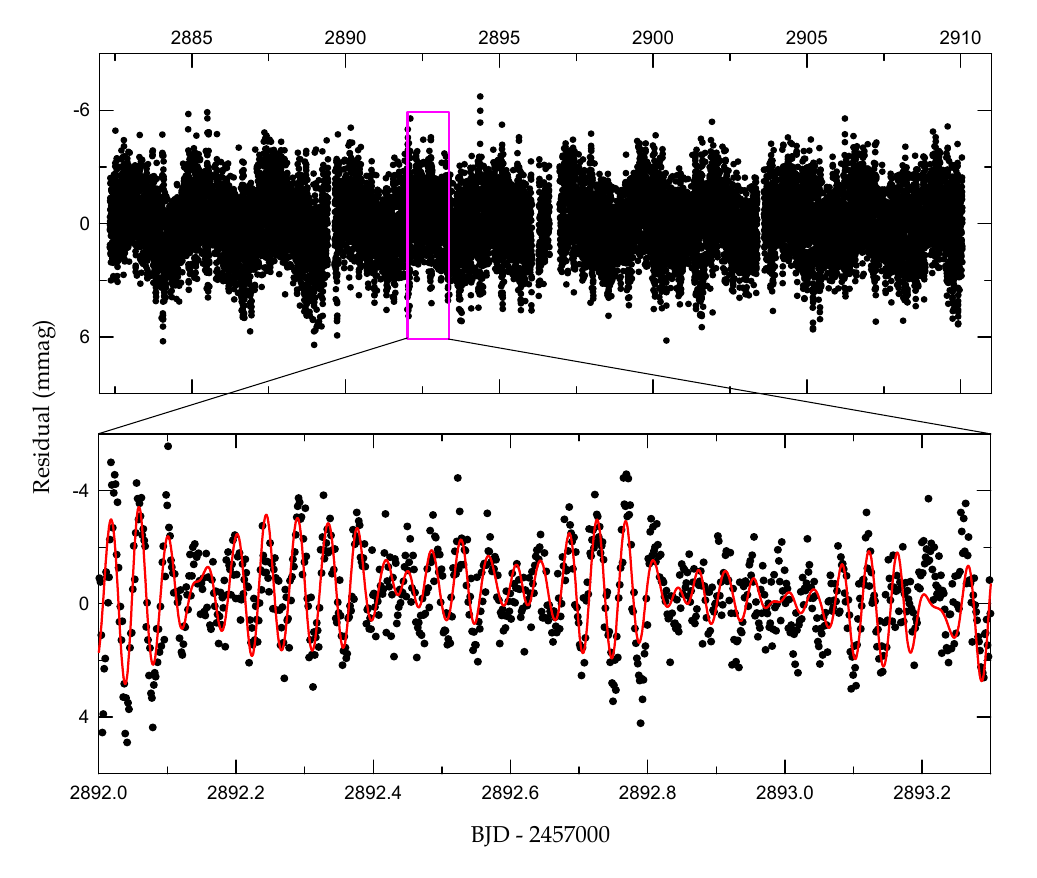}
\caption{Light curve residuals as a function of BJD. The lower panel provides a magnified view of the region indicated 
by the box in the upper panel. The red solid line represents the synthetic model curve based on a 15-frequency fit to 
the out-of-eclipse data. } 
\label{Fig6}
\end{figure}

For the reliable extraction and refinement of pulsation properties, we obtained the light-curve residuals 
through cycle-by-cycle binary modeling \citep{Lee2016}. To this end, we divided the TESS S58 light curve into 12 segments 
according to the orbital cycle. Because the final segment contained too few measurements for a reliable fit, it was combined 
with the preceding segment, resulting in 11 segments for individual analysis. For each segment, we fixed the binary parameters 
to the values listed in Table~\ref{Tab3}, except for the reference epoch ($T_0$) and the primary luminosity ($L_{\rm A}$), 
which were adjusted to account for possible cycle-to-cycle shifts in eclipse timings and variations in the light level, 
respectively. We then subtracted the corresponding best-fitting binary model from each segment and combined the residuals 
(Figure~\ref{Fig6}). 

The frequency analysis was conducted within the \texttt{PERIOD04} framework \citep{Lenz+2005}. To mitigate the impact of 
signal dilution during eclipses, only the out-of-eclipse data (within phase ranges of 0.044--0.456 and 0.544--0.956) were used 
in the analysis. We restricted the frequency search to the Nyquist limit of 360 $\text{day}^{-1}$, corresponding to 
the TESS sampling rate. The pre-whitening procedure was iteratively performed until no significant periodicities remained in 
the residuals \citep{Lee+2014}. Following the criteria suggested by \citet{Baran+2021}, only peaks with S/N exceeding 5.0 were 
considered statistically significant. The noise levels were computed over a 5 day$^{-1}$ window centered on each frequency in 
the Fourier spectrum. 

\begin{figure}
\includegraphics[]{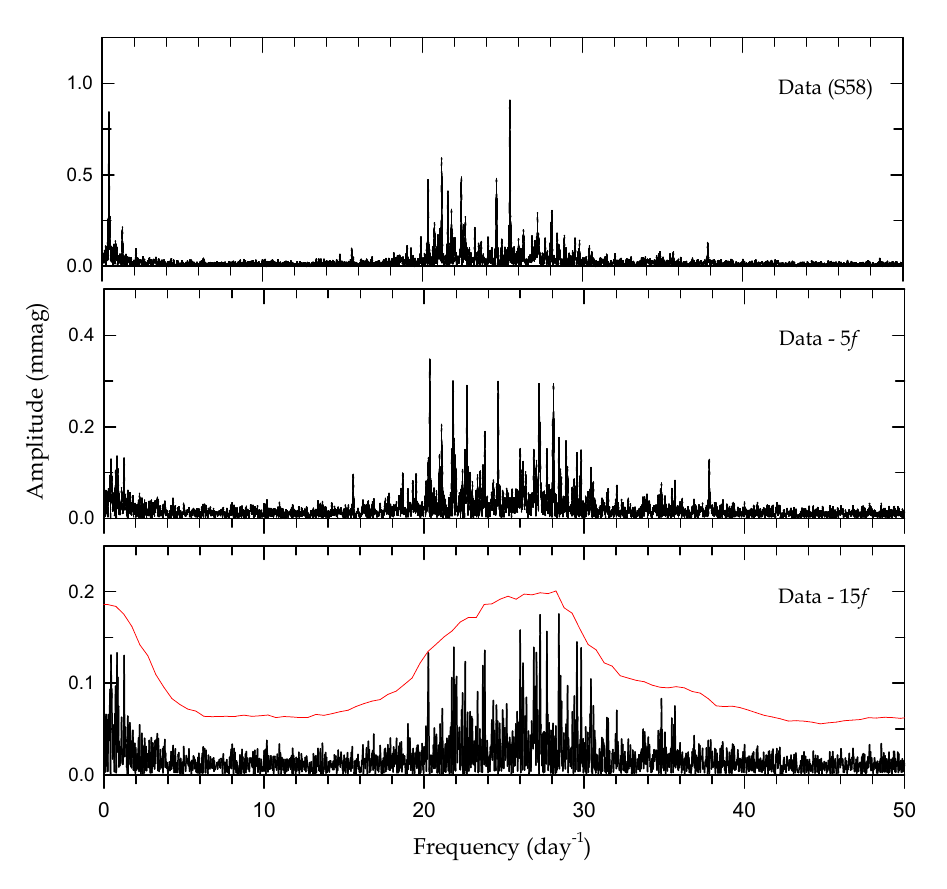}
\caption{\texttt{PERIOD04} periodograms for the outside-eclipse TESS residuals. The middle and bottom panels display 
the amplitude spectra after the successive pre-whitening of the first five frequencies and all 15 frequencies, respectively. 
The red solid line in the bottom panel corresponds to five times the noise spectrum. }
\label{Fig7}
\end{figure}

Through this iterative process, a total of 15 significant frequencies were successfully extracted. Their derived parameters, 
with uncertainties estimated via the formalisms of \citet{Kallinger+2008}, are summarized in Table~\ref{Tab5}. Based on 
these results, the synthetic curve calculated using \texttt{PERIOD04} is overplotted as a solid line on the residuals in 
the lower panel of Figure~\ref{Fig6}. The corresponding amplitude spectra are displayed in Figure~\ref{Fig7}, where the middle 
and bottom panels illustrate the pre-whitening process for the first five frequencies and the final 15 frequencies, respectively. 
To identify intrinsic pulsations and distinguish them from signals potentially associated with orbital effects or 
imperfect binary modeling, we investigated harmonics and linear combinations among the detected signals ($f_i$) and 
the orbital frequency ($f_{\rm orb}$), such as $f_i \pm Nf_{\rm orb}$ and $Mf_i \pm Lf_j$ (where $L, M$, and $N$ are integers). 
For this analysis, we adopted the Rayleigh criterion of $f_R = 0.036 \text{ day}^{-1}$ as the frequency resolution. As listed 
in the fourth column of Table~\ref{Tab5}, a substantial number of frequencies appear to be related to 
the orbital frequency or other detected frequencies through various combinations. 

For comparison, the results from \citet{Pawar+2024} are presented in the fifth and sixth columns of Table~\ref{Tab5}. 
They identified seven frequencies with S/N $> 5.0$ by analyzing the full TESS S58 residuals after removing their binary model. 
This count is markedly lower than the 15 frequencies recovered in our study. While the origin of this discrepancy is unclear, 
it may be partially attributed to differences in the residual light curves used for analysis. Specifically, our study excluded 
the eclipse intervals to ensure signal purity, whereas \citet{Pawar+2024} incorporated data from all orbital phases. 
With the exception of $f_2$ (corresponding to their $f'_6$), which is likely related to the orbital frequency, 
their identified frequencies agree with ours within 0.0153 day$^{-1}$. This difference is well within our Rayleigh resolution, 
underscoring the consistency between the two studies regarding these frequencies. 

The frequencies in Table~\ref{Tab5} are listed in the order of detection. When $f_3$ and $f_4$ were first extracted, 
$f_3$ had a larger amplitude than $f_4$, whereas their amplitude ordering reversed as additional frequencies were included 
in the fit. Because the final amplitude difference is smaller than the combined uncertainty, the amplitudes alone do not 
establish which frequency is the parent. We therefore list $f_3 \simeq f_4-3f_{\rm orb}$ and $f_4 \simeq f_3+3f_{\rm orb}$ 
as alternative possible combination identifications, with the parent frequency remaining uncertain. 

If $f_3$ is an independent pulsation mode, the frequency ratio $f_3/f_1$ = 0.832 is close to the theoretical value 
($\sim$0.845) expected for the second (2H) and third (3H) radial overtone modes in $\delta$ Sct stars \citep{Breger1979}. 
We calculated the pulsation constants for the primary component using the period–mean density relation given by 
$Q_i$ = $f_i^{-1}$$\sqrt{\rho_{\rm A} / \rho_\odot}$. The resulting values, $Q_1$ = 0.017 days and $Q_3$ = 0.020 days, 
are consistent with the third and second radial overtone $p$-modes, respectively \citep{Fitch1981}. However, if $f_4$ is 
instead the parent frequency, the ratio $f_4/f_1$ = 0.881 does not support the same overtone identification. 
The small observed amplitudes of $f_1$ and $f_3$, both below 0.9 mmag, also warrant caution in interpreting 
them as radial modes. Low-amplitude peaks in $\delta$ Sct stars can arise from higher-degree non-radial modes, as discussed 
for FG Vir by \citet{Daszynska-Daszkiewicz+2006}. Nevertheless, radial modes with amplitudes comparable to or smaller 
than those of non-radial modes have also been identified in $\delta$ Sct stars \citep{Breger2000}, so the observed amplitudes 
alone do not establish the mode geometry. We therefore regard the proposed radial-mode identification as tentative.

\section{Discussion and Conclusion} \label{sec_conclusion}

In this study, we have characterized the fundamental and pulsation properties of the PEB system HIP 7666 by integrating 
ground-based echelle spectroscopy from BOES and MRES with space-based TESS photometry. The double-lined RVs were extracted 
using the broadening function technique, and the atmospheric parameters of the primary component were determined to be 
$T_{\rm eff,A} = 7376 \pm 72$ K and $v_{\rm A}\sin i = 56.6 \pm 5.3$ km s$^{-1}$. Our observed $v_{\rm A}\sin i$ is slightly 
higher than the calculated synchronous velocity ($v_{\rm sync}\sin i \approx 45.4$ km s$^{-1}$), suggesting that the primary 
may be in a marginally supersynchronous state. 

The simultaneous modeling of the spectroscopic and photometric data shows that HIP 7666 is a dEB in a circular orbit. 
The fundamental parameters of both components were determined with high precision: the masses and radii to within 2 \%, 
and the luminosities to within 5 \%. For the primary component, we obtain $M_{\rm A} = 1.822 \pm 0.029$ $M_\odot$, 
$R_{\rm A} = 2.154 \pm 0.017$ $R_\odot$, and $L_{\rm A} = 12.28 \pm 0.52$ $L_\odot$. For the secondary component, 
the derived values are $M_{\rm B} = 1.246 \pm 0.020$ $M_\odot$, $R_{\rm B} = 1.170 \pm 0.015$ $R_\odot$, and 
$L_{\rm B} = 1.89 \pm 0.09$ $L_\odot$. These results are based on an RV dataset that is significantly more extensive than 
that of \citet{Feng+2021}. Consequently, our revised mass ratio ($q=0.684$) and the resulting absolute parameters provide 
a more robust and self-consistent description of the system’s physical and evolutionary properties. The excellent agreement 
between our EB-based distance ($316 \pm 8$ pc) and the Gaia DR3 parallax-based distance ($313 \pm 1$ pc) serves as 
independent validation of our binary solution. 

\begin{figure}
\includegraphics[scale=1.0]{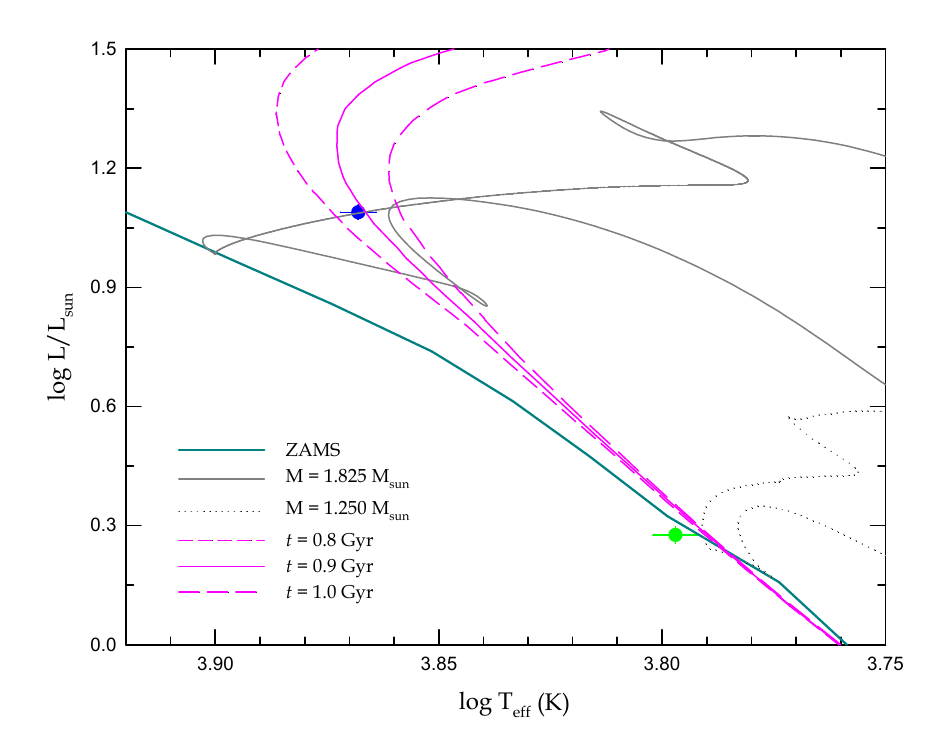}
\caption{H-R diagram for HIP 7666 A (blue circle) and B (green circle) compared with PARSEC models adopting the best-fit 
metallicity of $Z$ = 0.03 \citep{Bressan+2012}. The gray solid and dotted lines represent the evolutionary tracks for 
masses of $M$ = 1.825 $M_\odot$ and $M$ = 1.250 $M_\odot$, respectively. The zero-age main sequence (ZAMS) is indicated 
by a cyan solid line, while three isochrones ($t$ = 0.8, 0.9, and 1.0 Gyr) are plotted as pink short-dashed, solid, and 
long-dashed lines. } 
\label{Fig8}
\end{figure}

The precisely determined fundamental parameters enable a direct comparison with stellar evolutionary models. To assess 
the evolutionary status of HIP 7666, we plotted the individual components on the H-R diagram (Figure~\ref{Fig8}) and compared 
their positions with PARSEC v1.2S stellar models\footnote{\url{https://stev.oapd.inaf.it/PARSEC/}} \citep{Bressan+2012}. 
An initial comparison using the atmospheric parameters derived for solar metallicity favored a supersolar metallicity of $Z = 0.03$ 
and an age of approximately 1.0 Gyr. We therefore refined the atmospheric parameters of the primary by adopting this metallicity. 
The resulting stellar parameters support $Z = 0.03$ and yield a best-fit age of $t = 0.9$ Gyr. In Figure~\ref{Fig8}, the solid and 
dotted gray lines represent the evolutionary tracks for 1.825 $M_\odot$ and 1.250 $M_\odot$, corresponding to the primary and 
secondary masses, respectively. The H-R diagram indicates that the $\delta$ Sct pulsating primary is on the main sequence (MS), 
while its companion is located close to the zero-age main sequence (ZAMS). 

Furthermore, to verify the system’s population membership, we computed the Galactic kinematics as $(U, V, W)\footnote{Positive 
toward the Galactic center, Galactic rotation, and North Galactic Pole.} = (-22.88 \pm 0.76, 210.23 \pm 0.74, -6.46 \pm 0.32)$ 
km s$^{-1}$, based on the Gaia DR3 astrometric measurements and our derived systemic velocity, following the procedure of 
\citet{Lee+2024}. In the $U-V$ plane, HIP 7666 lies within the region occupied by the thin-disk population as characterized 
by \citet{Pauli+2006}. This kinematic classification is consistent with the relatively young age and supersolar metallicity 
($Z = 0.03$) inferred for the system. Our frequency analysis of the TESS light-curve residuals revealed 15 significant signals, 
most of which lie in the high-frequency range characteristic of $\delta$ Sct pulsations. Overall, HIP 7666 is a well-characterized, 
supersolar-metallicity PEB system that serves as a valuable benchmark for testing stellar evolution and pulsation models.

\begin{acknowledgements} 
We thank the anonymous referee for constructive comments and suggestions that helped improve the manuscript. 
This study uses echelle spectra obtained at BOAO and TNO, together with 120-s cadence light curves from the TESS mission. 
The TNO observations were carried out using the MRES spectrograph under Program ID TNTC7-002. 
\end{acknowledgements}

\clearpage 

\clearpage 
\begin{deluxetable}{lccclcc}
\tablewidth{0pt}
\tablecaption{TESS Eclipse Timings for HIP 7666 \label{Tab1} }
\tablehead{
\colhead{BJD}    & \colhead{Error} & \colhead{Min$\dag$} & \colhead{BJD}    & \colhead{Error} & \colhead{Min$\dag$} 
}
\startdata
2,459,882.56198  & $\pm$0.00005    & I             & 2,459,897.98167  & $\pm$0.00014    & II           \\
2,459,883.74825  & $\pm$0.00009    & II            & 2,459,899.16750  & $\pm$0.00006    & I            \\
2,459,884.93425  & $\pm$0.00004    & I             & 2,459,900.35387  & $\pm$0.00008    & II           \\
2,459,886.12058  & $\pm$0.00013    & II            & 2,459,901.53979  & $\pm$0.00011    & I            \\
2,459,887.30635  & $\pm$0.00006    & I             & 2,459,902.72614  & $\pm$0.00009    & II           \\
2,459,888.49260  & $\pm$0.00014    & II            & 2,459,903.91214  & $\pm$0.00009    & I            \\
2,459,889.67859  & $\pm$0.00013    & I             & 2,459,905.09820  & $\pm$0.00010    & II           \\
2,459,890.86498  & $\pm$0.00009    & II            & 2,459,906.28416  & $\pm$0.00007    & I            \\
2,459,892.05096  & $\pm$0.00014    & I             & 2,459,907.47049  & $\pm$0.00009    & II           \\
2,459,893.23716  & $\pm$0.00009    & II            & 2,459,908.65634  & $\pm$0.00006    & I            \\
2,459,894.42310  & $\pm$0.00013    & I             & 2,459,909.84286  & $\pm$0.00015    & II           \\
2,459,895.60924  & $\pm$0.00008    & II            &                                        
\enddata
\tablenotetext{\dag}{Type of eclipse. }
\end{deluxetable}

\begin{deluxetable}{lrrc} 
%\tabletypesize{\tiny}  %\small
\tablewidth{0pt}                    
\tablecaption{Radial velocities of HIP 7666$\dag$ \label{Tab2}}                                                                            
\tablehead{ 
\colhead{BJD}     & \colhead{$V_{\rm A}$}    & \colhead{$V_{\rm B}$}    & \colhead{Spectrograph} \\ [-1.0ex]
                  & \colhead{(km s$^{-1}$)}  & \colhead{(km s$^{-1}$)}  &                          
}
\startdata                                                                                                                  
2,458,158.9307    & $-41.95\pm0.42$          & $  81.77\pm1.29$         & BOAO/BOES              \\
2,458,425.9218    & $ 30.37\pm0.46$          & $ -25.79\pm0.95$         & BOAO/BOES              \\
2,458,425.9429    & $ 23.71\pm0.49$          & $ -27.89\pm1.08$         & BOAO/BOES              \\
2,458,425.9640    & $ 28.47\pm0.47$          & $ -12.34\pm0.53$         & BOAO/BOES              \\
2,458,452.9141    & $-72.48\pm0.43$          & $ 124.29\pm1.09$         & BOAO/BOES              \\
2,458,452.9353    & $-68.98\pm0.43$          & $ 119.67\pm1.09$         & BOAO/BOES              \\
2,458,452.9564    & $-66.49\pm0.41$          & $ 115.52\pm1.11$         & BOAO/BOES              \\
2,458,452.9776    & $-62.41\pm0.43$          & $ 109.43\pm1.15$         & BOAO/BOES              \\
2,458,452.9987    & $-59.11\pm0.41$          & $ 105.19\pm1.21$         & BOAO/BOES              \\
2,458,453.0199    & $-55.44\pm0.42$          & $  99.13\pm1.11$         & BOAO/BOES              \\
\enddata 
\tablecomments{This table is available in its entirety in machine-readable form. A portion is shown here for guidance regarding its form and content.}
\tablenotetext{\dag}{$V_{\rm A}$ and $V_{\rm B}$ represent the measured RVs of the primary and secondary stars, respectively. }
\end{deluxetable}

\begin{deluxetable}{lcc}
\tablewidth{0pt} 
\tablecaption{Light and RV Parameters of HIP 7666 \label{Tab3}}
\tablehead{
\colhead{Parameter}               & \colhead{Primary (A)} & \colhead{Secondary (B)}                                                  
}                                                                                     
\startdata 
$T_0$ (BJD)                       & \multicolumn{2}{c}{2,459,892.050858$\pm$0.000082}    \\
$P_{\rm orb}$ (day)               & \multicolumn{2}{c}{2.37222122$\pm$0.00000033}        \\
$a$ ($R_\odot$)                   & \multicolumn{2}{c}{10.872$\pm$0.054}                 \\
$\gamma$ (km s$^{-1}$)            & \multicolumn{2}{c}{7.64$\pm$0.45}                    \\
$K$ (km s$^{-1}$)                 & 93.16$\pm$0.70         & 136.17$\pm$0.90             \\
$q$                               & \multicolumn{2}{c}{0.6842$\pm$0.0068}                \\
$i$ (deg)                         & \multicolumn{2}{c}{81.366$\pm$0.053}                 \\
$T_{\rm eff}$ (K)                 & 7376$\pm$72            & 6272$\pm$66                 \\
$\Omega$                          & 5.756$\pm$0.028        & 7.522$\pm$0.035             \\
$\Omega_{\rm in}$$\rm ^a$         & \multicolumn{2}{c}{3.215}                            \\
$X$, $Y$                          & 0.641, 0.259           & 0.645, 0.227                \\
$x$, $y$                          & 0.448$\pm$0.056, 0.284 & 0.648$\pm$0.050, 0.261      \\
$l/(l_{\rm A}+l_{\rm B}+l_3)$     & 0.8190$\pm$0.0031      & 0.1435                      \\
$l_3$$\rm ^b$                     & \multicolumn{2}{c}{0.0375$\pm$0.0024}                \\
$r$ (pole)                        & 0.1967$\pm$0.0012      & 0.1074$\pm$0.0012           \\
$r$ (point)                       & 0.1998$\pm$0.0012      & 0.1079$\pm$0.0013           \\
$r$ (side)                        & 0.1980$\pm$0.0012      & 0.1076$\pm$0.0012           \\
$r$ (back)                        & 0.1994$\pm$0.0012      & 0.1079$\pm$0.0013           \\
$r$ (volume)$\rm ^c$              & 0.1981$\pm$0.0012      & 0.1076$\pm$0.0013           \\ [1.0mm]
\enddata                                                                                
\tablenotetext{\rm ^a}{Potential for the inner critical Roche surface.}
\tablenotetext{\rm ^b}{Value at 0.25 orbital phase. }
\tablenotetext{\rm ^c}{Mean volume radius.}
\end{deluxetable}

\begin{deluxetable}{lcc}
\tablewidth{0pt} 
\tablecaption{Absolute Parameters of HIP 7666 \label{Tab4}}
\tablehead{
\colhead{Parameter}           & \colhead{Primary (A)} & \colhead{Secondary (B)}                                                  
}                                                                                                                                     
\startdata                                                                                                                            
$M$ ($M_\odot$)               & 1.822$\pm$0.029       & 1.246$\pm$0.020           \\
$R$ ($R_\odot$)               & 2.154$\pm$0.017       & 1.170$\pm$0.015           \\
$\log$ $g$ (cgs)              & 4.032$\pm$0.006       & 4.397$\pm$0.011           \\
$\rho$ ($\rho_\odot$)         & 0.183$\pm$0.005       & 0.781$\pm$0.032           \\
$v_{\rm sync}$ (km s$^{-1}$)  & 45.92$\pm$0.36        & 24.94$\pm$0.33            \\
$v$$\sin$$i$ (km s$^{-1}$)    & 56.6$\pm$5.3          & \,                        \\
$T_{\rm eff}$ (K)             & 7376$\pm$72           & 6272$\pm$66               \\
$L$ ($L_\odot$)               & 12.28$\pm$0.52        & 1.89$\pm$0.09             \\
$M_{\rm bol}$ (mag)           & 2.01$\pm$0.05         & 4.04$\pm$0.05             \\
BC (mag)                      & 0.04$\pm$0.01         & $-$0.01$\pm$0.01          \\
$M_{\rm V}$ (mag)             & 1.97$\pm$0.05         & 4.05$\pm$0.05             \\
Distance (pc)                 & \multicolumn{2}{c}{316$\pm$8}                     \\
\enddata
\end{deluxetable}

\begin{deluxetable}{lcccclccc}
%\begin{deluxetable}{lc>{\color{red}}ccclccc}
\tablewidth{0pt}
\tablecaption{Multi-frequency Solution for HIP 7666 \label{Tab5}}
\tablehead{
\multicolumn{4}{c}{This Work}                                         && \multicolumn{2}{c}{\citet{Pawar+2024}}          && \colhead{Difference}   \\ [0.5mm] \cline{1-4} \cline{6-7} \\[-3.0ex]
\colhead{Frequency}    & \colhead{Amplitude} & \colhead{S/N} & \colhead{Remark}       && \colhead{Frequency}     & \colhead{Amplitude}   && \colhead{$f_i-f'_j$}                       
}                                                                                                                                                                              
\startdata                                                                                                                                                               
$f_{1}=25.4656(5)$     & 0.865(66)           & 22.43           &                        &&  $f'_{1}=25.4701(12)$   & 0.911(19)             && $-0.0045$              \\
$f_{2}= 0.4203(6)$     & 0.741(63)           & 20.17           & $f_{\rm orb}$          &&  $f'_{6}=0.4551(27)$    & 0.400(20)             && $-0.0348$              \\
$f_{3}=21.1939(7)$     & 0.536(51)           & 17.84           & $f_4-3f_{\rm orb}$ (?) &&  $f'_{3}=21.2031(19)$   & 0.567(19)             && $-0.0092$              \\
$f_{4}=22.4278(7)$     & 0.601(58)           & 17.78           & $f_3+3f_{\rm orb}$ (?) &&  $f'_{2}=22.4233(18)$   & 0.627(20)             && $+0.0045$              \\
$f_{5}=21.5836(7)$     & 0.519(53)           & 16.64           & $f_3+f_{\rm orb}$      &&  $f'_{4}=21.5832(22)$   & 0.504(20)             && $+0.0004$              \\
$f_{6}=20.3533(9)$     & 0.363(47)           & 13.28           & $f_3-2f_{\rm orb}$     &&  $f'_{7}=20.3380(27)$   & 0.407(20)             && $+0.0153$              \\
$f_{7}=24.6340(16)$    & 0.297(65)           &  7.83           & $f_1-2f_{\rm orb}$     &&                         &                       &&                        \\
$f_{8}=28.0866(18)$    & 0.270(69)           &  6.74           & $2f_7-f_3$             &&                         &                       &&                        \\
$f_{9}=21.8091(12)$    & 0.328(54)           & 10.38           &                        &&                         &                       &&                        \\
$f_{10}=27.1883(15)$   & 0.334(68)           &  8.34           & $f_1+4f_{\rm orb}$     &&  $f'_{5}=27.2003(22)$   & 0.494(20)             && $-0.0120$              \\
$f_{11}=22.6767(14)$   & 0.300(59)           &  8.70           & $f_9+2f_{\rm orb}$     &&                         &                       &&                        \\
$f_{12}=28.0289(23)$   & 0.218(68)           &  5.46           & $f_{10}+2f_{\rm orb}$  &&                         &                       &&                        \\
$f_{13}=21.0911(19)$   & 0.194(50)           &  6.64           &                        &&                         &                       &&                        \\
$f_{14}=20.9630(19)$   & 0.188(50)           &  6.47           & $f_9-2f_{\rm orb}$     &&                         &                       &&                        \\
$f_{15}=37.8159(17)$   & 0.123(28)           &  7.41           &                        &&                        &                        &&                        \\
\enddata
\tablecomments{Frequencies and amplitudes are given in units of day$^{-1}$ and mmag, respectively, and are listed in the order of detection. 
The parenthesized numbers represent the uncertainties in the last significant digit, calculated following \citet{Kallinger+2008}. 
The remarks for $f_{3}$ and $f_{4}$ indicate alternative possible combination identifications; the parent frequency remains uncertain. }
\end{deluxetable}

\end{document}